\documentclass[preprint]{aa}
\usepackage{natbib}
\bibpunct{(}{)}{;}{a}{}{,} 
\usepackage{graphicx}
\usepackage{txfonts}
\usepackage{umoline}
\usepackage{changes}

\def\msun{M$_{\odot}$}
\def\rsun{R$_{\odot}$}

\def\porb{P$_{\mathrm {orb}}$}
\def\mdot{$\dot M$}

\def\it{\sl}
\def\degs{\ifmmode ^{\circ}\else$^{\circ}$\fi}
\def\amin{\ifmmode ^{\prime}\else$^{\prime}$\fi}
\def\asec{\ifmmode ^{\prime\prime}\else$^{\prime\prime}$\fi}
\def\fd{\hbox{$.\!\!^{\rm d}$}}            

\def\degs{\ifmmode ^{\circ}\else$^{\circ}$\fi}
\def\amin{\ifmmode ^{\prime}\else$^{\prime}$\fi}

\def\eqalign#1{\null\,\vcenter{\openup1\jot \m@th
   \ialign{\strut\hfil$\displaystyle{##}$&$\displaystyle{{}##}$\hfil
   \crcr#1\crcr}}\,}
\usepackage{color}

\def\te{$T_{\mathrm {eff}}$}

\newcommand{\sdss}{V479\,And}

\def\apgt{\ {\raise-.5ex\hbox{$\buildrel>\over\sim$}}\ }
\def\aplt{\ {\raise-.5ex\hbox{$\buildrel<\over\sim$}}\ }

\titlerunning{Multiwavelength observations of V479 Andromedae}
\authorrunning{Diego Gonz\'alez et al.}

\begin{document}

\title{Multiwavelength observations of V479 Andromedae: \\ a close compact binary with an identity crisis.
\thanks{Photometry and reduced spectra are only available at the CDS via anonymous ftp to cdsarc.u-strasbg.fr (130.79.128.5) or via http://cdsarc.u-strasbg.fr/viz-bin/qcat?J/A+A/}}

\offprints{Diego Gonz\'alez--Buitrago \\  \email{dgonzalez@astrosen.unam.mx}}

\author{Diego Gonz\'alez--Buitrago,\inst{1}
 Gagik Tovmassian,\inst{1}
 Sergey Zharikov,\inst{1}
 Lev Yungelson,\inst{2}
\\ Takamitsu Miyaji,\inst{1}
Juan Echevarr\'{\i}a,\inst{3}
 Andres Aviles\inst{1}
\and Gennady Valyavin\inst{1}
}

\institute{
Instituto de Astronom\'{\i}a, Universidad Nacional Aut\'onoma
de M\'exico, Apartado Postal 877, Ensenada, Baja California, 22800 M\'exico,
dgonzalez, gag, zhar@astrosen.unam.mx
\and {Institute of Astronomy of the Russian Academy of Sciences, 48 Pyatnitskaya Str., 119017 Moscow, Russia.}
\and
Instituto de Astronom\'{\i}a, Universidad Nacional Autonoma de M\'exico,  Apartado Postal 70-264,
Cuidad Universitaria, M\'exico D.F., 04510  M\'exico,
}
\abstract
{}
{ We conducted a multi-wavelength study to unveil the properties of the extremely long-period cataclysmic variable \sdss.}
{We performed series of observations, including moderate to high spectral resolution optical spectrophotometry, X-ray  observations with {\sl Swift}, linear polarimetry and near-IR photometry.}
{ This  binary system is  a low-inclination $\sim 17^o$  system  with a 0.594093(4) day orbital period. The absorption line complex in the spectra indicate  a G8--K0 spectral type for the donor  star,  which has departed from the zero-age main sequence. This  implies a distance to the object of about 4 kpc.  The primary is probably a massive 1.1-1.4 M$_\odot$ magnetic white dwarf, accreting  matter at a rate \mdot$\, > 10^{-10}$\,\msun\, yr$^{-1}$. This rate can be achieved if the donor star fills its corresponding Roche lobe, but there is little observational evidence for a mass-transfer stream in this system. An alternative explanation is a stellar wind  from the donor star, although such a high rate mass loss is not anticipated from a subgiant.  If the  strongly magnetic white dwarf in \sdss\  is confirmed by future observations, the system  the polar with the longest observed orbital period. We also discuss the evolutionary state of \sdss. }
{}


\maketitle

\section{Introduction}

\sdss\ was identified as a cataclysmic variable by \citet{2005AJ....129.2386S}
in the Sloan Digital Survey (SDSS\,J001856.93+3454) because it shows narrow Balmer emission lines, He\,{\sc i},  and strong He\,{\sc ii}. The
two-hour follow-up observations   were not sufficient to detect any significant radial velocity variations to reveal an orbital period.
\citet{2005AJ....129.2386S} also performed polarimetry and found no  linear polarization within those two hours.
Similarly, \citet{2008MNRAS.386.1568D} found no orbital period of this system  from four nights of photometric observations.
\citet{2010arXiv1009.5803G} conducted an extensive  spectroscopic campaign in 2008-09  to pin down the orbital period. A $14.52\pm0.53$ hour period was
found by cross-correlating absorption features in the spectra of the object, with a K4\,V standard-star template.
Calculations of the orbital period, derived from the emission line radial velocity measurements  (fitted with a single Gaussian), did not match
the orbital period determined from  the absorption lines. Therefore, the authors assumed
that the object might be an asynchronous polar.
In this paper  we present new  extensive optical, IR, UV and X-ray data  and a comprehensive study of this object.

\begin{table}
\caption{Log  of time-resolved  observations.}
\begin{tabular}{l|cccll} \hline \\[0.1pt]
&  \multicolumn{4}{c}{Spectroscopy / 2.1m}   &         \\[0.1pt]
\hline \\[0.1pt]
Date  & Exp.  &   FWHM & Range &   N of        & Total      \\
           &     sec         &         \AA                         &       \AA          &            spectra                       &   hours \\[1pt]   \hline \\[0.1pt]
24/12/03$^*$   &  5400 & 3 & 3800-9200 &   1    &        \\
06/11/08    & 1200    & 4.1 & 3900-5950 &  22 &   8        \\
07/11/08     & 1200 & 4.1   & 3900-5950 &  19     & 7      \\
08/11/08     & 1200 & 4.1   & 3900-5950  & 25     & 9      \\
09/11/08     & 1200 & 4.1  & 3900-5950   &  10     & 4       \\
11/11/08    &  1200  &  4.1 &  6050-8100 &  10  &  4 \\
06/12/08        & 1200 &  4.1 & 3825-5875    & 22 &   8        \\
07/12/08     & 1200 & 4.1 & 3850-5900 & 9     & 3      \\
28/08/09     & 1200 & 4.1 & 3825-5875 & 17     & 7      \\
29/08/09     & 1200 & 4.1 & 3825-5875 & 10     & 4       \\
29/09/09     & 1200 & 4.1 & 3850-5900 &  22     & 8       \\
07/09/10     & 1200 &  2.1     & 4550- 5900 & 6       &  2     \\
08/09/10     & 1200 &  2.1    & 4550- 5900 & 18       &  8     \\
05/10/10     & 1200 &  2.1     & 4550- 5900 & 8       &  3     \\
07/10/10     & 1200 &   2.1    & 4550- 5900 & 12       &  6     \\[1pt]
\hline
\hline \\[0.1pt]
&\multicolumn{4}{c}{UV, optical and IR photometry }   &           \\[1pt]
\hline  \\
Date  & Exp.  &   Filter & Telescope &   N of        & Total      \\
           &     sec         &            &     Instrument           &        images                           &   hours \\ [1pt]   \hline \\[0.1pt]
$^\star$ & 176   &  FUV  & {\em GALEX} & 2  &  \\
$^\star$ & 176   &  NUV & {\em GALEX} & 2  &  \\
26/11/10$^\dagger$   & 553-1691 & UVM2 & {SWIFT} & 8 & 11\\
27/11/10 $^\dagger$  & 927-1636 &   UVW1 & {SWIFT} & 11 & 16 \\
06/09/10    &  300        & I & 1.5m/RUCA & 52  &    6      \\
07/09/10    &  300        & I & 1.5m/RUCA & 60  &   7        \\
08/09/10    & 300      &  I   & 1.5m/RUCA & 42   & 6     \\
09/09/10    & 300    &   I    &  1.5m/RUCA & 45  & 6     \\
05/10/10    & 300    & I & 1.5m/RUCA &  67 &   7       \\
06/10/10    & 30    & I & 1.5m/RUCA &  100 &   4       \\
07/10/10   & 300    & I & 1.5m/RUCA &  93  &   9        \\
29/10/10   & 480    & J & 1.5m/Camila&  35 &   4  \\
30/10/10   & 480   & J  & 1.5m/Camila &  53   & 6    \\
31/10/10   & 480   & J  & 1.5m/Camila &   51 &    6  \\[1pt]
\hline
\hline \\[0.1pt]
&\multicolumn{4}{c}{Linear polarimetry / 0.84m}   &           \\[1pt]
\hline \\[0.1pt]
Date  & Exp.  &   Filter & Telescope &   N of        & Total      \\
           &     sec         &            &    Instrument            &        images                           &   hours \\ [1pt]   \hline \\[0.1pt]
30/09/10   & 300    & V & POLIMA &  67 &   7       \\
01/10/10   & 300    & V & POLIMA &  43 &   4       \\
30/10/10   & 300    & V & POLIMA &  64  &   6        \\
31/10/10   & 300    & V & POLIMA &  60 &   6       \\
01/11/10   & 300    & V & POLIMA &  43  &   4        \\

\hline
\\
\end{tabular}
\begin{tabular}{l}
$^*$ Observed by {\em SDSS}. \\
$^\star$ {Observed by {\em GALEX}. ObjID \# 6372252849676485379} \\
$^\dagger$ {Observed by {\em SWIFT/UVOTA} OBS\_ID \# 00031872 } \\
\end{tabular}
\label{tab:log}
\end{table}

\section[]{Observations}
\subsection{Spectroscopy.}
Time-resolved spectroscopy of \sdss \,  was performed with the 2.1 m telescope at the Observatorio Astron\'omico Nacional\footnote{http: www.astrossp.unam.mx}
at San Pedro M\'artir, Baja California, M\'exico (OAN SPM) in 2008 and 2009  with the Boller \& Chivens spectrograph, using a 600 grooves mm$^{-1}$ grating
and a $24\,\mu m \, 1024\times1024$\,pixel SITe CCD  with a spectral resolution of 4.1\,\AA. In  2010 we made
two additional runs, using the 1200 grating with a $15\,\mu m \, 2048\times2048$\,pixel Thomson CCD, with a spectral resolution of 1.8\,\AA. 
The wavelength calibration was made with an arc lamp taken every tenth exposure, and spectrograph flexion with azimuth was also corrected using for night-sky lines.  The spectra of the object  were flux-calibrated using spectrophotometric standard stars observed 
during the same night. The instrument cannot automatically rotate the slit to the corresponding parallactic angle,  and for simplicity we  routinely used an E-W slit orientation. In addition, the slit width was kept narrow ($180\,\mu$m = 2 arc sec) for better resolution. These two factors make a correct flux calibration 
difficult, although this is usually not  a problem in radial velocity (RV) studies.  However, on the night of September 07, 2010 we observed the object and several late spectral type standards  using a wide $450\,\mu$m slit to ensure better flux calibration.  The standard long-slit reduction of the data was made using IRAF\footnote{IRAF is distributed by the National Optical Astronomy Observatory, which is operated by the Association of Universities for Research in Astronomy (AURA) under cooperative agreement with the National Science Foundation.}
procedures after applying basic CCD preliminary procedures. Only cleaning cosmic rays, which are abundant on 1200 sec exposures,  was made with the external task {\it lacos} \citep{2001PASP..113.1420V}. The log of spectroscopic observations is given in Table~\ref{tab:log}.

We also used the SDSS spectrum of the object from DR8. The original spectrum of \sdss\ was obtained on December 10, 2003, 
but the new reduction was made with an improved calibration using the software version v5-3-12 
of February, 2008. The SDSS spectrum  covers a wider range of wavelengths than our own spectra and provides a reliable
flux calibration.  
A list of the main emission lines and their measurements are presented in Table\,\ref{tab:emlines}.

\subsection{Photometry and photopolarimetry}

Time-resolved I and J bands near-IR photometry  and V-band linear polarimetry were obtained using the 1.5\,m and the 0.84\,m telescopes, respectively, at the OAN SPM.
The first was obtained using the direct CCD image mode
with the RUCA and CAMILA instruments$^1$, while the latter was obtained with the  POLIMA$^1$,  using four polarization positions (0, 45, 90, and 135 degrees).
Two polarimetric standards, one of which  has zero polarization, were observed together with the object for  calibration purposes.
The log of photometric observations is also given in Table~\ref{tab:log}.
The data were reduced  with IRAF and a pipeline software for POLIMA\footnote{http://www.astrossp.unam.mx/\~\,sectec/ web/instrumentos/polima/otrospolima.html}. The images were corrected for bias and flat-fields  before aperture photometry.  Flux calibration was performed using  standard stars from the lists of \citet{1992AJ....104..340L} and \citet{1996A&AS..119..547V}, observed in the same nights.
The  errors in optical V and I bands are estimated to be less than 0.05 magnitudes. We had problems with the telescope during  the IR J-band observations, and thus only a rough $\pm0.3$\,mag estimate of brightness is available. 

\subsection{UV and X-ray observations}

\sdss\  was observed with {\sl Swift} as a  target of opportunity (ToO target ID: 31872) for 31 hours on November 26 and  27, 2010 with a total on-source exposure time of 26.7 ks.
We used two  of the three instruments on board, of the {\sl Swift} gamma-ray burst explorer (see \citet{2004ApJ...611.1005G}):  the X-ray Telescope (XRT; e.g. \citet{2005SSRv..120..165B}) and the Ultraviolet/Optical Telescope (UVOT; e.g. \citet{2005SSRv..120...95R}).  
Standard data processing was made later at the  Swift Science Data Centre  in 2010 December. The results presented in this paper are based on data collected when the XRT was operating in the PC mode, in which full imaging and spectroscopic resolutions are retained, but timing resolution is limited to 2.5s.

Simultaneous UV images in the UVW1 and UVM2 filters, centered at 2600 and 2246 \AA, respectively, were obtained with the UVOT.  One entire binary orbital cycle was covered.  
Eleven images were collected with the UVW1 filter and eight with the UVM2 filter.
We used standard aperture  photometry to measure fluxes on calibrated images, which were supplied after pipeline reduction of the UVOT data. 

\begin{table*}
 \centering
    \caption{Measurements of the main emission lines }
\begin{tabular}{lcccccc  |  llcccc} \hline
&  \multicolumn{2}{c}{Wavelength}   &         &        &                      &    &            &  \multicolumn{2}{c}{Wavelength}   &   & &     \\
ID  &  air    &   measured  & Flux/2.0$\times10^{-14}$   & EW  & FWHM    &    &   ID  &  air    &   measured  & Flux/2.0$\times10^{-14}$   & EW  & FWHM   \\
    &  \AA    &    \AA      &    H$_\beta$   & \AA &   km/sec           &   &               &  \AA    &    \AA      &    H$_\beta$   & \AA &   km/sec\\
            \hline
            		 &               &      &       &	    &      &    &	        	 &               &       &           &	       &          \\
  H  {\sc i}     & 4101.73	 &   4101.76	&  0.64     &	 -30.6  &  713.9    &    & C  {\sc iii}   & 4640.03	 &   4641.49	&  0.04     &	 -2.2  &  639.3     \\
        *         &               &  4099.78   &  1.17     &	 -36.4 &  1068.0    &    &		 *	 &               &     &       &	 &     \\
  H  {\sc i}     & 4340.46	 &   4340.91	&  0.64     &	 -32.9  &  657.3    &    &	  He  {\sc ii}   & 4685.71	 &   4686.26	&  0.47     &	 -24.1  &  632.5      \\
       *          &               &   4339.03  &  1.15     &	 -37.0 &  919.5     &    &		*         &.		 &   4685.91	&  0.90     &	   -23.0  &  574.3    \\
                 &               &     &       &	  &                                                      &    &                *	 &               &   4685.26   &  0.91     &	  -26.7 &  697.9  \\
  He  {\sc i}    & 4471.48	 &  4472.00	&  0.22     &	-11.5	&  530.6    &    &	 H  {\sc i} 	 & 4861.32	 &   4861.73	&  0.67     &	 -34.7  &  629.5      \\
       *          &               &   4470.91  &  0.46     &	-13.7	&  771.8    &    &	*         &	 	 &  4861.52	&  1.00     &	  -27.8  &  611.6      \\
                    &               &     &       &	  &                                                      &    &      *	 &               &   4861.073  &  1.11     &	  -32.5 &  728.2   \\
  He  {\sc ii}   & 4541.59	 &   4541.84	&  0.03     &	 -1.9  &  618.3     &    &	 He  {\sc ii}   & 5411.53	 &   5412.93	&  0.07     &	 -4.0  &  507.2  \\
       *          &               &   4541.88  &  0.06     &	 -1.73  &  687.0    &    &		 & 	 &  	&       &	  &     \\
  N  {\sc iii}   & 4634.12	 &   4634.3	&  0.01     &	 -1.3  &  603.3     &    &	 H  {\sc i}     & 6562.80	 &   6563.37	&  0.61     &	 -32.3  &  494.5  \\
	           		 &               &      &       &	    &      &    &	        	 &               &       &           &	       &          \\
\hline \hline
\end{tabular}
\label{tab:emlines}
\end{table*}

\section{System composition}
\subsection{Spectroscopic orbital period}
\label{sec:op}

\begin{figure}

   \includegraphics[width=9cm,  clip]{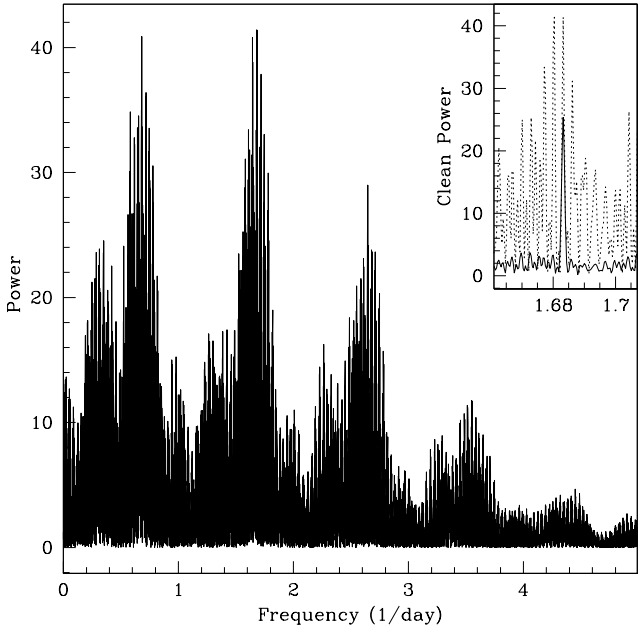}
  \caption{Power spectrum of the radial velocities of the secondary star. Peaks around 0.7 and 2.7 day$^{-1}$ are one-day aliases of the orbital period at 1.68325 day$^{-1}$. The aliases are strong because the orbital period is long and only part of a cycle is observed nightly.  In the inset box of the figure the {\it CLEAN}ed power spectrum peak is shown, marking the precise period of the system.  }
  \label{fig:power}
\end{figure}

The spectroscopic orbital periods of close binaries are better determined from the radial velocity variation of the absorption lines of the stellar components when available  than from the lines originating in the accretion disk.  In the case of \sdss\ we were able to use the absorption line complex around the $\lambda\lambda\,5050-5850$\,\AA\ region 
by a cross-correlating with a template spectrum of the corresponding class. The absorption features around 5200\,\AA\ are the strongest and  are very similar to
late-G or early-K stars.  We used the IRAF {\it xcsao} procedure for the cross-correlation. The regions that contained an emission line or were affected by a sky line falling 
into our selected range were excluded from the analysis.
The results presented here are based on a cross-correlation with the  spectrum of HD099491, a K0 IV star \citep{1976AJ.....81..245E}. 
We also used templates of  spectral types, varying from G8 to K2, with luminosity types V and IV, which do not show
notable differences between measured velocities.
Standard spectra of late stars observed simultaneously with the object  show similar results.

The radial velocities  of all spectra taken in 2008 to 2010 were measured. The velocities for 2008 and 2009 match those reported in \citet{2010arXiv1009.5803G}, where a K4\,V template was used. The RV were first analyzed for  periods using 
{\it CLEAN} \citep{1978A&A....65..345S}, a discrete Fourier transform method, which convolves the power spectrum with the spectral window to eliminate alias 
frequencies originating from uneven data distribution.  We  also conducted a Scargle-Lomb \citep{1982ApJ...263..835S} test on the data, 
which is a least-squares spectral analysis  similar to the Fourier one,  
 but which mitigates the  long-periodic noise in the long gapped records. 
The results of both period analyses are shown on Figure\,\ref{fig:power}. The Scargle-Lomb produces two  equal peaks at frequencies 1.6803 and 1.6832 day$^{-1}$, while
{\it CLEAN} gives a clear preference to the latter frequency, which we used hereafter as the orbital period of the system,  although the other period gives a similar RV curve 
and similar scatter of points to the fitted  sinusoidal curve. This period is within the errors of the value reported by \citet{2010arXiv1009.5803G}. 
The ephemerides, according to the combined 2008-2010 data, are
$$ {\rm HJD}= 2\,454\,776.3479(6) + 0\fd594093(4) \times {\rm E},  $$
where $\mathrm{HJD}_0$ at E=0 refers to the inferior conjunction of the secondary star. 

\subsection{Orbital parameters of the binary components}

\begin{figure}
 \includegraphics[width=9cm,  clip]{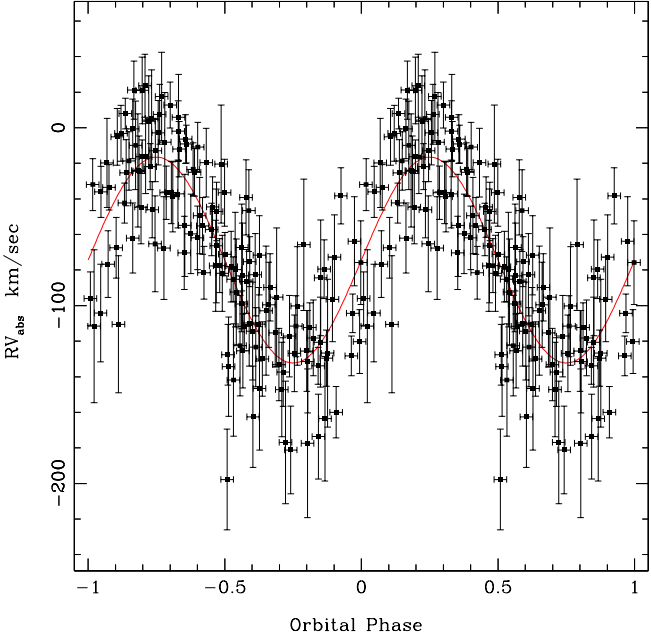}
  \caption{Radial velocities of the absorption line complex  determined by cross-correlating  the object with a standard K0 star template and folded with  the orbital period. The red line is the best-fit sinusoid to the measurements. }
  \label{fig:rvabs}
\end{figure}

The orbital parameters of the secondary star can be found from a sinusoidal fit of the form
$$RV(t) = \gamma+K_{\mathrm d}\times \sin(2\pi(t- \mathrm {HJD_0})/P)$$  to the RV curve.
A least-squares fit yields $K_{\mathrm d} = 59$\,km\,s$^{-1}$ and $\gamma = -73$~km\,s$^{-1}$,
where $P$ and $ \mathrm {HJD}_{ 0}$  have been taken from the calculated ephemeris. Figure\,\ref{fig:rvabs} shows the velocities folded with 
these parameters  together with the sinusoidal fit (Table\,\ref{tab:rvs}).

\begin{figure}
 \includegraphics[width=8.6cm,  clip]{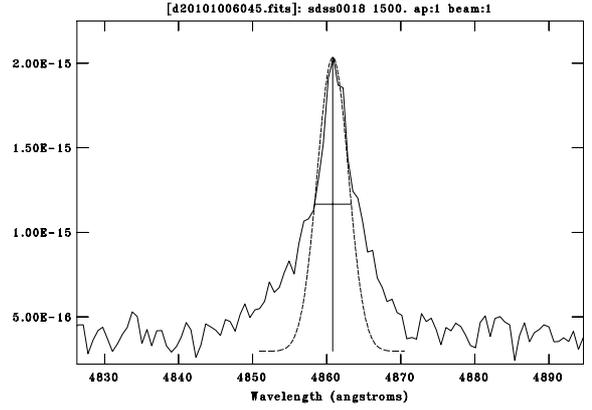}
  \caption{ Typical H$_\beta$ line profile from a spectrum obtained in 2010 with a 2.1 FWHM spectral resolution. A single-Gaussian profile with 5.0\AA\ FWHM is over-plotted on the line to demonstrate that the base of the line deviates from a Gaussian.  }
  \label{fig:profile}
\end{figure}

\begin{figure}

   \includegraphics[width=9cm, clip]{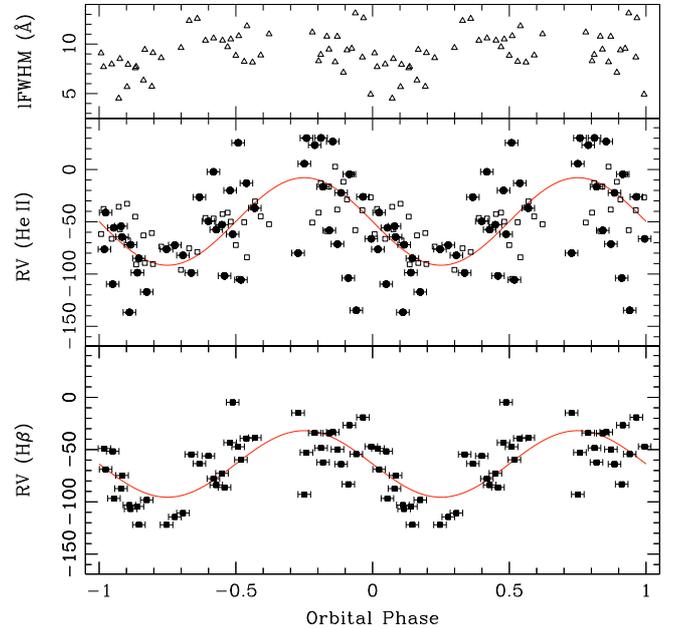}
  \caption{Emission line measurements of \sdss\ from 2010 higher resolution spectra. The radial velocities of H$_\beta$\ and  He\,{\sc ii}  measured by a double-Gaussian method are marked by solid black squares in the bottom  and middle panel. The horizontal bars attached to each point indicate the duration of individual exposures. The radial velocity curves obtained as sinusoidal fits to these points are overplotted. In the middle panel  measurements of He\,{\sc ii}  line by single-Lorentzian fits are marked by open squares. The upper panel shows measurements of the width of the Lorentzian profiles fitted to the He\,{\sc ii} line.}
  \label{fig:rvhem}
\end{figure}

The emission lines of this object are very intense and single-peaked. 
The average FWHM of  Balmer lines is 500-600 km\,s$^{-1}$, which is quite narrow for an ordinary CV. 
\citet{2010arXiv1009.5803G} measured the radial velocities of  these lines by fitting  a single Gaussian to the line profiles. This resulted in a scatter of points, with no
clear periodic nature. The same results were achieved by cross-correlating  the emission lines with a synthetic line constructed with a Gaussian profile of 5\,\AA\ width.
The scattered measurements were incorrectly interpreted by \citet{2010arXiv1009.5803G}.
The new, higher resolution spectra obtained in 2010 show that the  lines do not have a Gaussian profile.  The base of the lines is wider for a Gaussian,  
and there might be a contribution from a broader, but less intense component. An example of a H$_\beta$ line profile is shown in Figure\,\ref{fig:profile} with a single Gaussian fitted to the peak of the line.  To measure 
the wings of the emission lines, a double-Gaussian method was proposed by \citet{1980ApJ...238..946S}. We followed here a prescription developed by \citet{1983ApJ...267..222S}. 
Best results  for the set of 2010 spectra were achieved with  Gaussian widths of 5.3 \AA\ and a separation of 16--17 \AA\  ($\approx1000$ km\,s$^{-1}$).  These solutions 
show an orbital modulation with the same period as that obtained  from the secondary star and are, as expected, 
in counter phase with the absorption lines. The semi-amplitudes and systemic velocities obtained for H\,{\sc i} and He\,{\sc ii}  $\lambda 4686$\,\AA,
with the orbital period and zero phase fixed, are also summarized in Table\,\ref{tab:rvs}.  
We remeasured the He\,{\sc ii}  $\lambda 4686$\,\AA\ line also  with a single-Lorentzian  fit.  Unlike a single-Gaussian fit, this  takes  into account the wings of the line. 
The orbital solution is  similar  to that obtained with the double-Gaussian fitting,  but with a lower semi-amplitude of the radial velocities.
To ensure that the RV variability in the wings of the emission lines was not influenced by the absorption features from the donor,  we repeated the analysis after subtracting  a standard K0\,V spectrum, observed at the same time and with the same instrument setup, scaled to a flux and Doppler-shifted to a velocity corresponding to a contributing companion star.  We note that in this fit,  the profile of emission lines becomes  twice as narrow  at about orbital phase $\phi_{\mathrm {orb}} = 0.1-0.2$ ,  as can be seen in the top panel of Figure\,\ref{fig:rvhem}.
In the two lower panels of Figure\,\ref{fig:rvhem} we show the measured RVs for H$_\beta$\ (bottom) and  He\,{\sc ii} $\lambda\, 4686$\,\AA\ (middle) folded with the orbital period and phase obtained from the analysis of absorption lines.  Corresponding fits are also shown as sinusoidal curves.The filled square symbols mark the RVs measured with the double-Gaussian method, while the open squares are measurements with the 
single-Lorentzian fits to He\,{\sc ii}  $\lambda 4686$\,\AA. The uppermost panel shows the distribution of the line widths of the Lorentzian fits.

\begin{figure*}[!ht]

   \includegraphics[width=9cm,  clip]{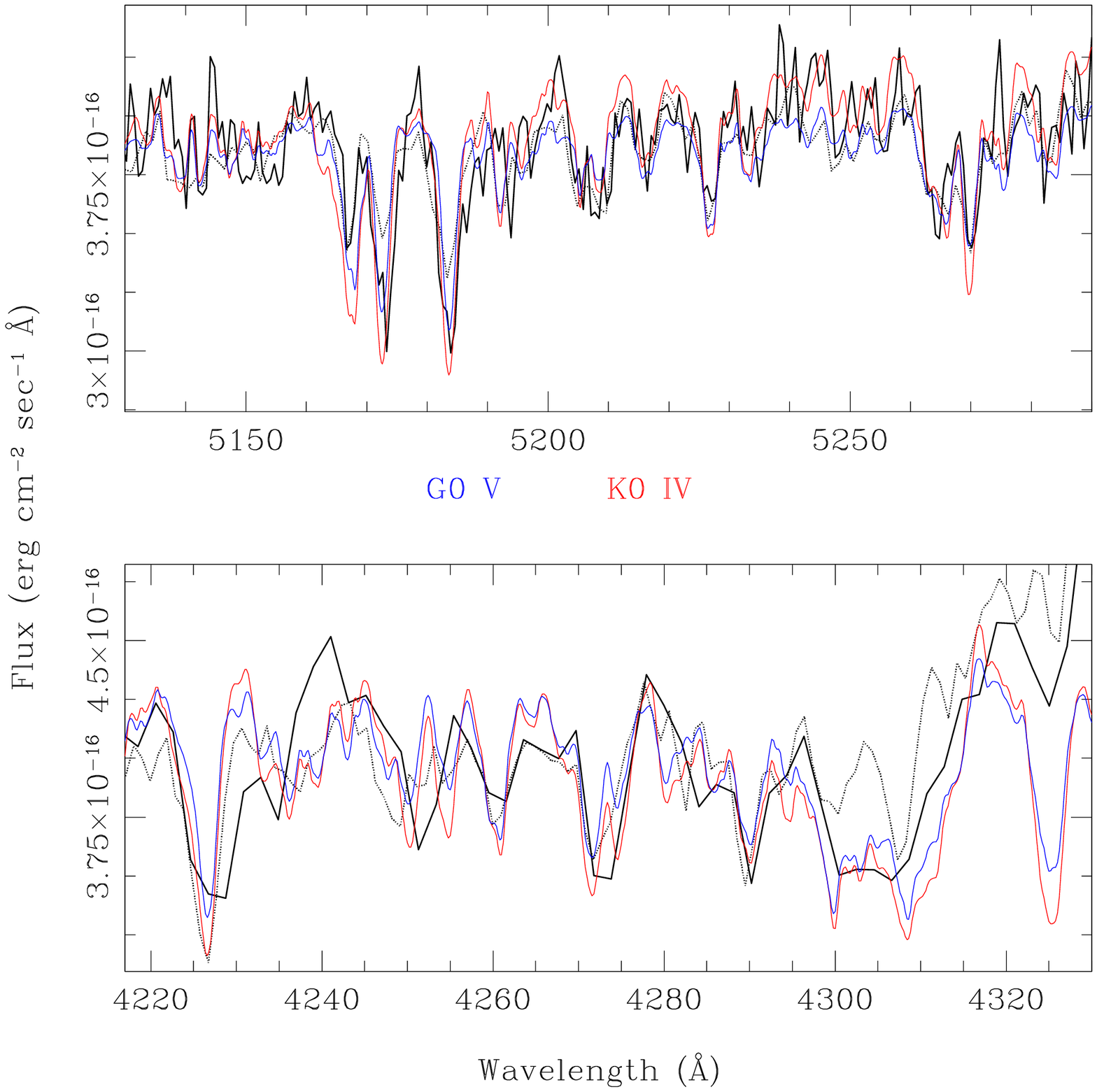}
    \includegraphics[width=9cm,  clip]{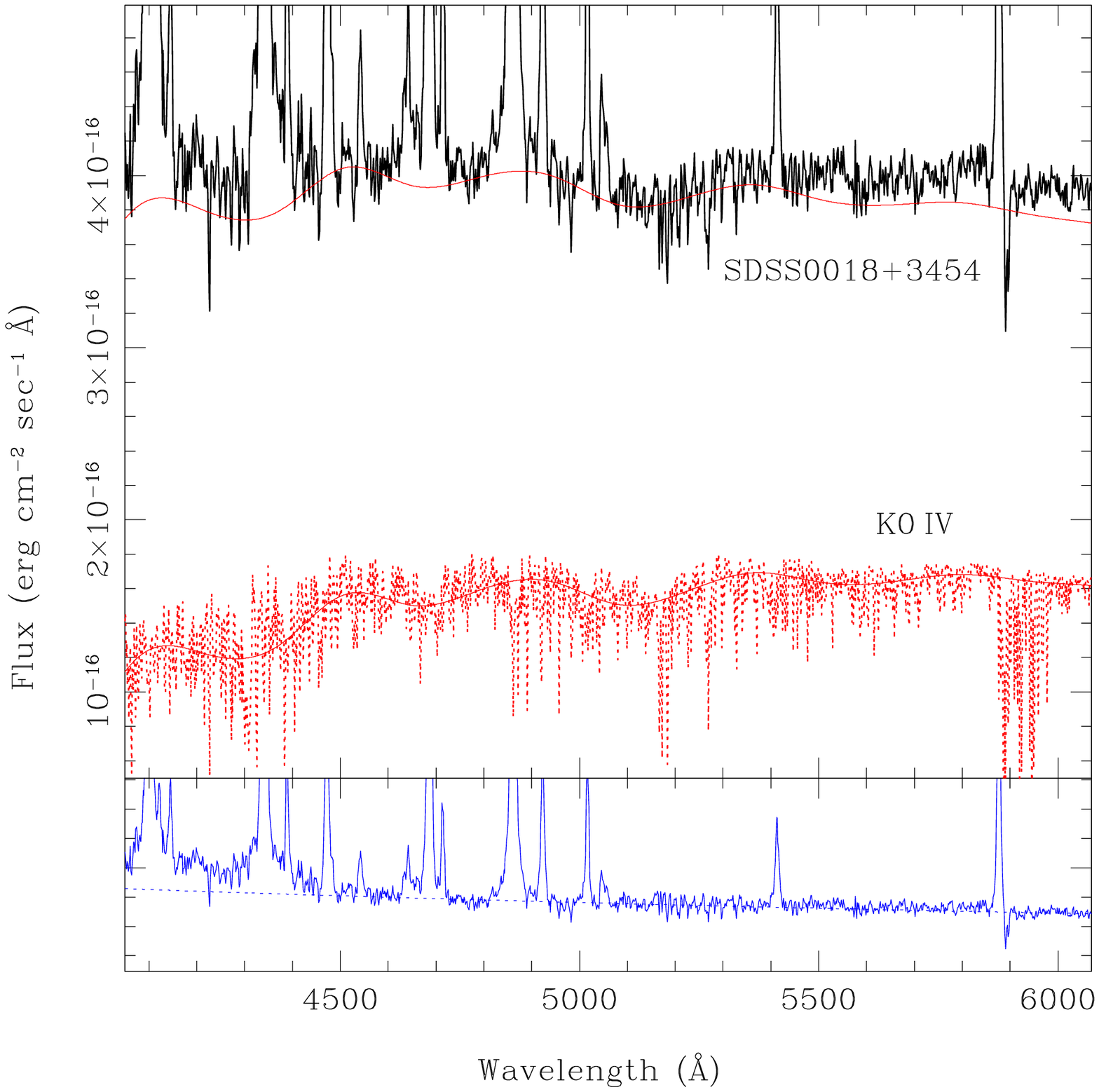}
   \caption{ Spectra  of \sdss\ (black)  and  standard stars. In the right upper panel we show the observed 
spectrum of the object  along with a K0\,IV star (red, dotted), scaled to reflect the  real contribution of the secondary star to the total flux.  The continuous red line over the K0 star is a high-order polynomial fit  to the spectrum excluding strong absorption features. In the lower right panel, the residual (\sdss\  minus K0\,IV) is plotted. It is well-described by a power law plotted by the dashed (blue) line.  This power law  is summed with the fit to the secondary star spectrum  and plotted over the object in the upper panel to illustrate coincidences of the troughs in the continuum. 
In the left panels  zoomed portions of the spectra are displayed in the $\lambda\lambda\,4220-4340$\,\AA\  (bottom)  and $\lambda\lambda\,5100-5300$\,\AA\ (top)  intervals. The thick black line in  the left panels corresponds to  the SPM  spectrum, while the dotted line shows the SDSS spectrum. Two standard stars are plotted in colors,  G0\,V in blue, 
and K0\,IV in red. The standard star spectra are scaled  to overlap with  the object in each spectral interval separately.  The SPM spectra in the left panels  are combined  after  correction for the secondary star orbital motion to increase the signal-to-noise ratio.  }
  \label{fig:sp5}
\end{figure*}
\begin{table}
 \centering
    \caption{Radial velocity fit parameters.} 
\begin{tabular}{l|ccc} \hline
Line ID  &   $\gamma$  & Velocity &   Phase shift         \\ 
           &    km sec$^{-1}$         &     km sec$^{-1}$    &    relative to HJD$_0$  \\    \hline
abs. lines   & $-73.3\pm2.9$    & $58.9\pm4.0$ &  0.0$^\dagger$         \\
H$_\beta$  &   $-63.7\pm3.5$    &  $31.8\pm5.5$  &  0.49   \\
He\,{\sc ii}  $\lambda 4686$\,\AA   &   $-49.7\pm5.9$    &   $41.9\pm9.2$  &   0.51  \\
\hline
\end{tabular}
\begin{tabular}{l}
$^\dagger$ {Fixed;  $P_{orb}$ given by the ephemeris.} 
\end{tabular}
\label{tab:rvs}
\end{table}

It is not clear where the wings of the emission lines originate. 
Comparing   the relative intensity of the Balmer lines from Table\,\ref{tab:emlines}  with those computed by \citet{1980ApJ...235..939W} shows that  the Balmer decrement 
of \sdss\ is much flatter than  expected  from the accretion disk of a CV,  even with a very high $10^{-9}$\,\msun/year mass--transfer rate. The  strong He\,{\sc ii} line is not commonly observed in an ordinary accretion disk. Moreover,  He\,{\sc ii} spots in the disk or elsewhere in the binary would not have similar widths 
and  orbital phases  as the Balmer lines, which are normally produced by the entire disk.  
In polars,  the emission lines often contain a narrow component, which at times can be  intense and dominate the line profile, but this component is formed in a stream of 
matter and usually produces a high radial velocity component (HVC)  \citep[e.g.][]{1983ApJ...271..735S,1997A&A...319..894S, 1999ASPC..157..133T}.  A ballistic stream, or a coupling region should  produce a dominant high-velocity contribution,  and  it will not be exactly shifted 0.5  to the phase of the absorption lines. 
At the radius of magnetosphere, which  prevents the formation of a  disk, the speed in a ballistic stream will reach hundreds of km/sec, and even in a low-inclination system 
its radial velocity will easily surpass the orbital velocity of either stellar component.

Because we do not detect high radial velocities in \sdss, but  rather see  low radial velocities  exactly in  counter phase to the absorption lines, we conclude that  the emission lines
originate somewhere in the vicinity of the accreting star,  but probably do not arise from the usual components in 
a standard Roche-lobe-filling/stream-mass-transfer CV model.   To summarize, we consider the  orbital velocity  of the accreting component in this binary system to be within the range of  semi-amplitudes measured by He\,{\sc ii} and H$_\beta$\ lines.

\subsection{The spectral classification of the secondary star}
\label{sec:spcl}

The right panel of Figure\,\ref{fig:sp5} shows  portions of  \sdss\  spectrum   
together with   two late-type stars.  The starts of G0 to K2 spectral types with no peculiarities and the highest signal-to-noise ratio Êwere selected from ELODIE archive \citep{1996A&AS..119..373B} to compare them with the object. The spectra were degraded to a spectral resolution corresponding to our observations.  In the left panel, we show two spectral intervals: $\lambda\,4220-4340$\,\AA\  (bottom)  and $\lambda\,5100-5300$\,\AA\ (top) in which a number of absorption features belonging to a late-type star show up in the spectra of \sdss. The  complex of Fe\,II + Mg\,I  features around $\lambda\,5173\,\AA$ and Fe\,I~+ Ca\,I   around $\lambda\,5270\,\AA$ and others in the $\lambda\,5000-5800\,\AA$  range indicate the wide range of spectral types from G0 to K4.  But additional lines in a more contaminated   $\lambda\,4200-4340\,\AA$\  region,
particularly Ca\,I  (4226), Cr\,I (4254, 4290), and  Fe\,I (4271)   allow one to seek a balanced solution  between individual line depths and the spectral energy distribution.   
To successfully fit absorption lines in both selected regions,  the contribution from the continuum of the secondary star should be adequately taken into account.  In the left panels the standard spectra are arbitrarily scaled  to overlap with the absorption features of the object, i.e.,  the continuum of each comparison star is adjusted to the observed continuum in each region separately.  G-K type stars have numerous absorption lines that form troughs of different depth and width in their spectra. Their locations move with the change of the overall shape of the continuum depending on the temperature. We delineated these features  by fitting a high-order polynomial to the spectra of the comparison stars (excluding strong absorption lines) in a strictly homogeneous way.  
In the right upper panel of  Figure\,\ref{fig:sp5} we show an example of a standard spectrum of a K0\,IV star and the fit to it.  Obviously,  the troughs in the continuum of a late-type star are visible also in the spectrum of the object.
The depth of the feature around 5200\,\AA\ and the slope of the continuum between the H$\gamma$ and H$\delta$ lines in the observed spectrum can be  best reproduced by  summing of G8 -- K0\,IV and a power law, although it is very difficult to evaluate the agreement quantitatively. Employing  other spectral types will require  more complex  contribution than that of a simple power law  from the rest of the system.  

 The lower right panel shows the residual spectra after 
subtracting  the K0 IV star.  The overplotted dashed line represents a power law accounting for difference in the continuum of the observed object and a donor star.  
Subtracting  a standard earlier than G8 or later than K2 produces an odd continuum,  different than a smooth power law expected from  accretion-powered radiation. This shows that the secondary contributes  about 45\% of the flux to the total luminosity of the object at 5000\,\AA.

\subsection{Masses and radii}

\begin{figure}[t]
  \includegraphics[width=9cm,bb = 10 210 570 700, clip]{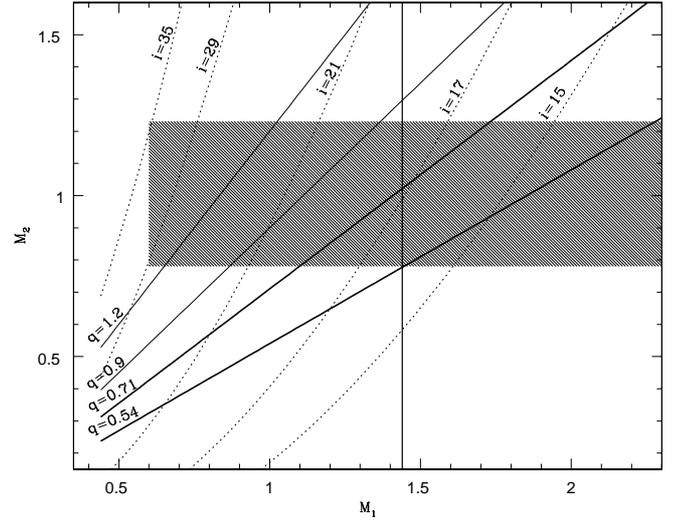}
  \caption{ $M_{2}$ - $M_{1}$ diagram for V479\,And. 
The  solution probably lies  between thick diagonal lines corresponding to $q=0.54\ {\rm {and}}\ 0.7$ as measured by RV amplitude ratio of He\,{\sc ii}  and  H$_\beta$\  emission lines to the absorption. Dashed curves denote the orbital inclination of the system. The shaded area corresponds to  the mass range of a K0 star from ZAMS to luminosity class IV.}
  \label{fig:mr}
\end{figure}

Obtaining mass and radius limits of the binary components may help us to understand the dynamics of the accretion region.
 We can determine $$q=M_2/M_1=K_1/K_2$$ from Table\,\ref{tab:emlines}. We obtain $q=0.54$  if we take the semi-amplitude of  the primary star  from 
the H$_\beta$ line and $q=0.71$  if the determination from the He\,{\sc ii}\,$\lambda 4686$\,\AA\  is correct.   The solution is somewhere in between since these these two values largely overlap if errors are taken into account. The $M_2 - M_1$ diagram is presented in Figure\,\ref{fig:mr}. Diagonal lines denote different mass ratios $q$, with upper and lower  limits corresponding to values determined from   H$_\beta$ and He\,{\sc ii} marked by thicker lines. Since the
semi-amplitude  depends on the inclination angle, the different inclination curves of the system are  plotted. 
The vertical line shows  the Chandrasekhar mass limit and divides the white dwarf and neutron star regimes.
The shaded area marks the mass range of a K0 star from ZAMS to luminosity class IV, limited  at the bottom by a 0.79\,\msun\  main-sequence K0 star and at the top by a 1.23\,\msun\,, K0\,IV star \citep{2010A&ARv..18...67T}; at the left it is limited by a $~0.6$\,\msun\ white dwarf mass (corresponding to the peak in the oxygen-carbon white dwarf mass  distribution).
Note that  our upper $q$ crosses the Chandrasekhar mass limit at about M$_2\approx 1.03$\,\msun.  Therefore, it is probable that the accretor in this system is a massive white dwarf or even a neutron star. 
The possible presence of a massive white dwarf in this system is an intriguing feature. 
Several exceptionally long-period systems appear to have massive white dwarfs, e.g. RU\,Peg with P$_{\mathrm {orb}} \sim 9$h; \citep{1990MNRAS.246..654F}, and EY Cyg with P$_{\mathrm {orb}} \sim11$h; \citep{2007A&A...462.1069E}. A disproportionally large portion of CVs with high-mass white dwarfs are also magnetic which, as \citet{1999ApJ...525..995V} pointed out, can be the result of  their peculiar formation. Finally, we can point out from Figure\,\ref{fig:mr} that the possible solutions show that we are dealing with  a very low inclination system. 

\begin{figure}[ht]
  \includegraphics[width=9cm, clip]{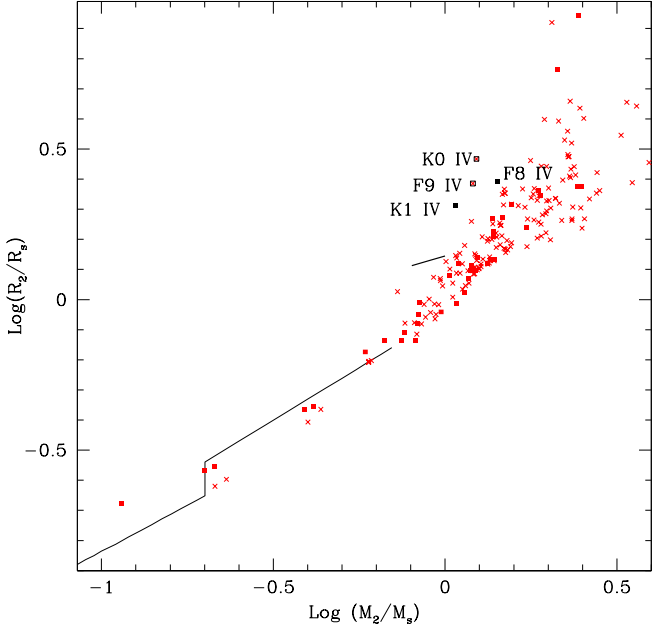}
  \caption{$M_{\mathrm d}$ / $R_{\mathrm d}$ relation for low-mass stars. Crosses are parameters of normal stars measured in detached binaries, while solid red squares are measurements obtained from interferometry. A few subgiants from the sample are marked individually. The possible location of \sdss\ is marked by a thick black line. The thin, broken black line corresponds to main-sequence donors in CVs with shorter orbital periods. }
  \label{fig:r2m2}
\end{figure}

The secondary star in \sdss\ cannot be a zero-age main-sequence K star since for an orbital period of 14.26 hours the star would be
far from filling its Roche lobe. More likely the companion is an  evolved star. 
To assess its luminosity class we can
calculate the Roche lobe size of the companion  for mass ratio values $0.54 \le q \le 0.7$ and a range of primary masses
1.1 - 1.4 \msun.  
These values are shown in Figure\,\ref{fig:r2m2}  with the short, thick black line. The  crosses mark very precise mass and radius determinations for  detached spectroscopic binaries and the filled squares those for well-known interferometric binaries  \citep{2010A&ARv..18...67T}.  Subgiant stars are singled out; open black squares correspond  to detached binaries, and solid black squares  to interferometric binaries. The broken thin black line is the mass-radius  relation for main-sequence donor stars of CVs with orbital periods
$< 6$\,hours  \citep{2006MNRAS.373..484K}. 
It is evident from Figure \ref{fig:r2m2} that  \sdss\ is certainly off the zero-age main sequence, although it is still far from reaching
the K subgiant region. The mean mass and radius values of \sdss,
obtained form the Roche lobe constraints discussed above, are  $M_2 = 0.79$ \msun\ and $R_2 = 1.29$ \rsun. 
We can compare these values with some of the closely similar spectroscopic binaries. For example, a comparison with HS Aur B, a K0V star 
with $M = 0.88$ \msun\ and $R = 0.87$ \rsun\ that is  close to the zero-age main sequence with a large surface gravity log g = 4.5 (cgs),
shows that its radius is much smaller than that of \sdss.
If we look at the subgiants AI\,Phe A, a K0IV with $M = 1.23$ \msun\ and $R = 2.93$ \rsun, 
and V432 Aur, an F9IV star with $M = 1.20$ \msun\ and $R = 2.43$ \rsun, both with low surface gravity values
log g = 3.59 (cgs) and log g = 3.75 (cgs), respectively (both plotted in the Figure \ref{fig:r2m2}), we observe that they have  radii  nearly a factor of three larger than \sdss.
Finally, we look at two intermediate surface gravity stars that have masses and radii comparable to \sdss, the G5~V stars
V568 Lyr A and RW Lac A,
both of which appear well off the zero-age main sequence, but have radii $1.2 - 1.4$~\rsun\  that are comparable to \sdss. 
  We must be very cautious in accessing the donor star parameters because intermediate to low-mass stellar evolution models  have been
highly criticized for being unable to predict the radii and effective temperatures of detached eclipsing binary stars. Recent models that incorporate magnetic fields help improve the situation \citep{2012ApJ...761...30F}, and we are convinced that the donor star in \sdss\, is magnetically active (see the following Sections). 

\subsection{Distance estimates}

We can derive rough distance estimates to the system. A firm lower limit on the 
distance would be $3.0\pm0.2$\,kpc, taken from the distance module and assuming the donor is a main-sequence star with a flux F$_\lambda\approx1.8\times10^{-16} {\mathrm {erg\,cm}}^{-2} {\mathrm {sec}}^{-1} {\AA}^{-1}$  at 5500\,\AA\  and an
interstellar extinction of E($B-V$)=0.056 in the direction of \sdss.  But the donor star is certainly not a zero-age main-sequence star. We have shown in the previous section that the  
Roche lobe size of  \sdss\ is probably about  1.3\,\rsun. This would imply a distance to the object
of about 4 kpc. This places the object in the halo of the Galaxy, at 1.8\,kpc above the Galactic plane.

\subsection{Photometric variability and polarimetry}

\begin{figure*}[!ht]

   \includegraphics[width=9cm, bb = 0 210 570 720, clip]{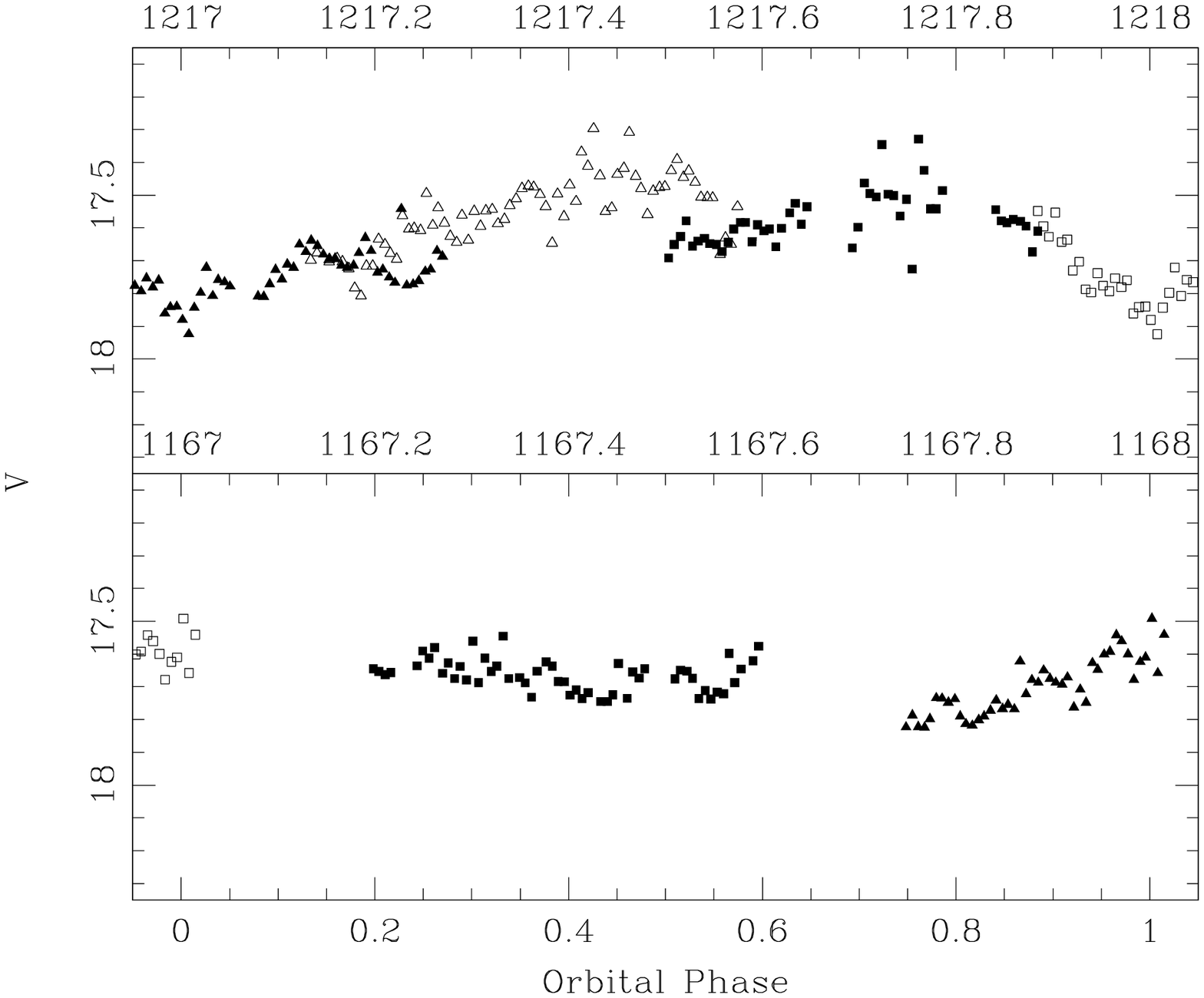}
    \includegraphics[width=9cm, bb = 0 210 570 720, clip]{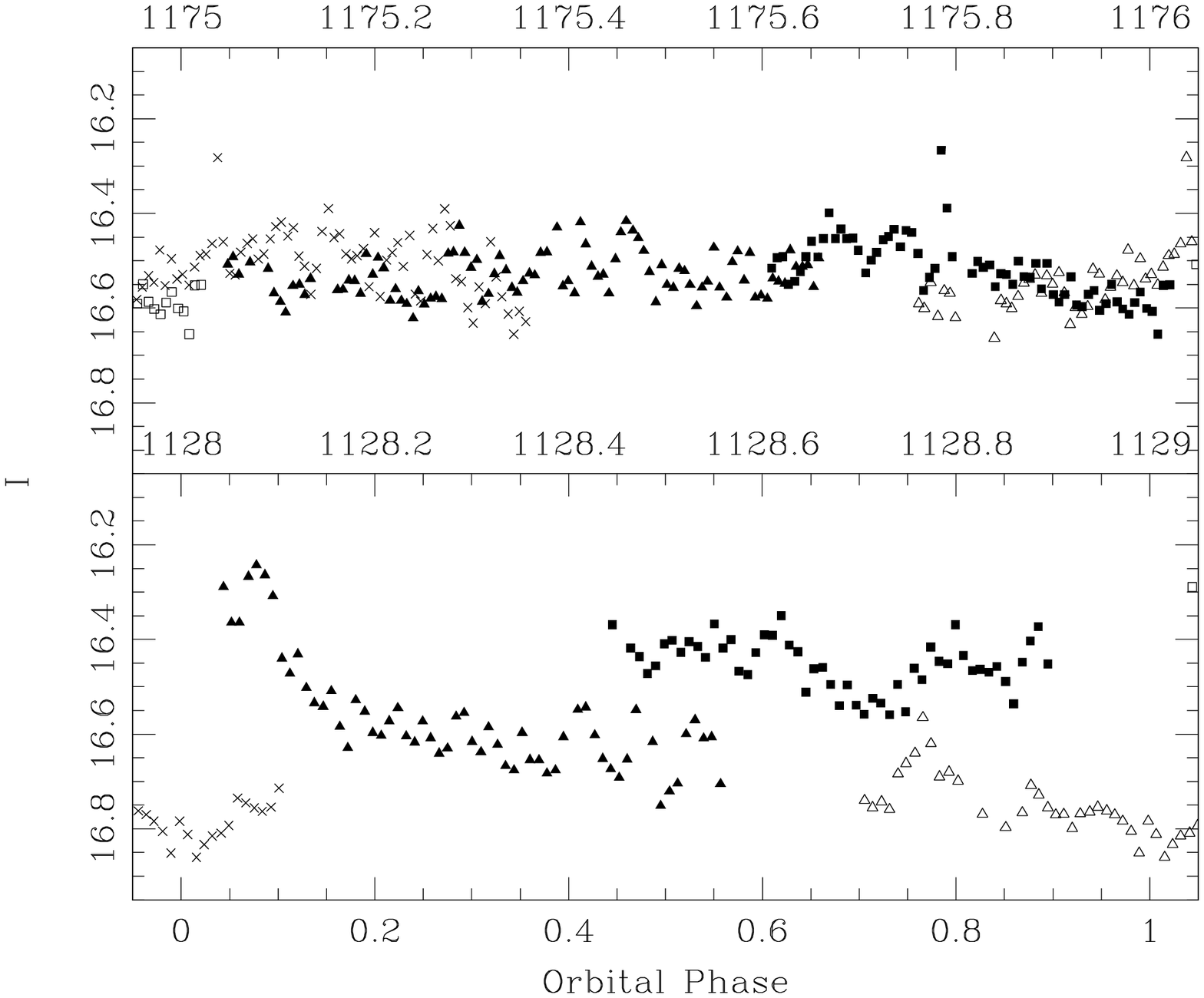}
  \caption{Light curves of \sdss\  in $V$ \& $I$ bands as indicated on the panels and obtained at four different epochs are  folded with the orbital period. The integer numbers at the X-axes above each panel correspond to the orbital cycle in which  the data plotted with filled squares were taken, the decimals correspond to the orbital phases. The data from immediately previous and following orbital cycles  are marked with  different symbols.  The magnitudes at the Y-axes are calibrated using secondary standards in the field of the object. }
  \label{fig:photVI}
\end{figure*}

The limited photometric coverage in the $V$ and $I$ bands shows variability up to  0.5 mag. There is no clear  evidence of an orbital modulation, i.e.,  the   pattern of variability does not repeat from  cycle to cycle. In  Figure\,\ref{fig:photVI}  the light curves of the object are presented, folded with the orbital period determined from spectroscopy.  The $V$ and $I$ magnitudes were obtained by observations of \citet{1992AJ....104..340L} stars and defining  secondary standards in the field of the object.  In each panel, in addition to the orbital phase marked at the bottom, the X-axis at the top of panels indicate the mean E$^{th}$ cycle given by the ephemerides (section\,\ref{sec:op}).  More than one orbital cycle has been observed at each observational run, thus the symbols refer to different orbital cycles around the mean. In the $V$-band, during cycle 1217, a variability appears that resembles  a double-humped light curve, typical of close binaries with irradiated secondaries. But  during the previous  cycle 1167, the light curve is  fairly flat.
Similarly, in the  $I$-band, the light curve around cycle 1175 is practically flat, while at the earlier
cycle 1128 there is a strong, irregular variability. This behavior is similar to that obtained from the  optical photometry  by  \citet{2008MNRAS.386.1568D}; this authors reported  irregular variations of light, but no clear orbital modulation.
The infrared $J$ data, obtained in SPM, are very noisy and cover only a fraction of the binary orbit. Our average flux estimate J=16.25(25) is significantly different from the 2MASS magnitude J = 15.77(8), H = 15.22(10), K = 15.29(16).  Evidently,  the IR flux is also variable.

We  detected no  linear polarization in the $V$ band above a 2\% threshold.   \citet{2005AJ....129.2386S} also failed to detect any polarization, although their orbital phase coverage is not complete. While a polarization detection is a good indicator of magnetic activity, linear polarization is not always observed in magnetic systems such as polars.

In the X-ray, we found a variability with a systematic pattern. Two orbital periods where covered with {\sl Swift},  during which the  light curve  shows two pulse-like events. With only two orbital cycles and the limited time-resolution
given by the {\sl Swift} orbits, it is not possible to determine the period. Nevertheless, folding the X-ray light curve with the orbital period determined from the spectroscopy  shows a  single-hump curve in which the pulses of  both cycles overlap.
The  X-ray light curve is presented in Figure\,\ref{fig:xuv} (bottom panel).  The data are folded  according to the spectroscopic ephemerides.
The UV magnitudes obtained in parallel with the X-ray data with the UVOT detector  show a similar behavior, confirming that the X-ray pulses are not accidental spikes. The UV data  taken in two different bands in each cycle are  shown in the
same figure (upper panel). The light curve in filter $UVW1$  clearly shows a  trend similar to that of the X-ray data and is modulated with the orbital period. However, the phase of the peak in the near-UV trails the X-ray pulse by $\sim0.1$\  orbital phase. In  the $UVM2$ filter the data corresponding to the peak in $UVW1$ are missing.

\subsection{X-ray spectrum}
\label{sec:xrayspec}
  
\begin{figure}[!hb]
  \includegraphics[width=9cm, bb = 0 150 570 720, clip]{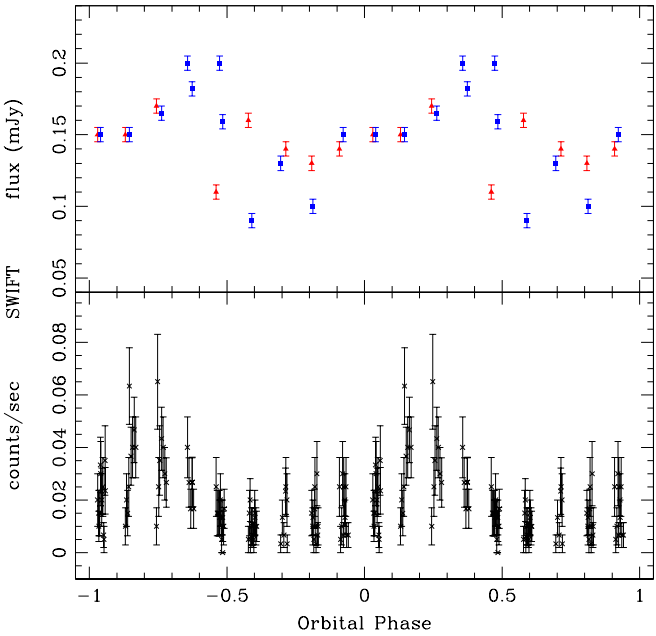}
  \caption{X-ray and UV light curves folded with the orbital period and ephemerides determined from the spectroscopy. The two orbital periods were observed by {\sl Swift}. The X-ray flux peaks briefly at around orbital phase 0.2, while there seems to be a deep minimum at the phase between phases 0.5 and 0.7.  The UV light curve in two filters (UVW1 blue squares and UVM2 red triangles) presented in the upper panel shows a behavior similar to that in the X-ray, except that the maximum peak is shifted relative to X-rays by 0.2 orbital phases and reaches maximum at 0.4. }
  \label{fig:xuv}
\end{figure}

 The X-ray spectrum  from the 26.7 ks of {\it Swfit} XRT observations accumulated from a 9.8$^{\prime\prime}$ 
radius circle around the source contains only  $\sim380$ photons.  Thus, it was meaningless to perform an in-depth spectral analysis.  Most of these photons arrived during the 
 ``peak phase'', which corresponds to the orbital phases from roughly 0.12 to 0.30 (see Figure \ref{fig:xuv}). 
The peak phase has a much softer spectrum than the non-peak phase and varied component,
 suggesting a compact eclipsed emission with a lower temperature than the non-eclipsed diffuse
 gas component.  
 A much higher signal-to-noise, phase-resolved spectroscopy with {\it XMM-Newton} and/or {\it Suzaku} would be very useful for understanding this object.

\subsection{Spectral energy distribution}
\label{sec:sed}

\begin{figure}
  \includegraphics[width=9cm,  clip]{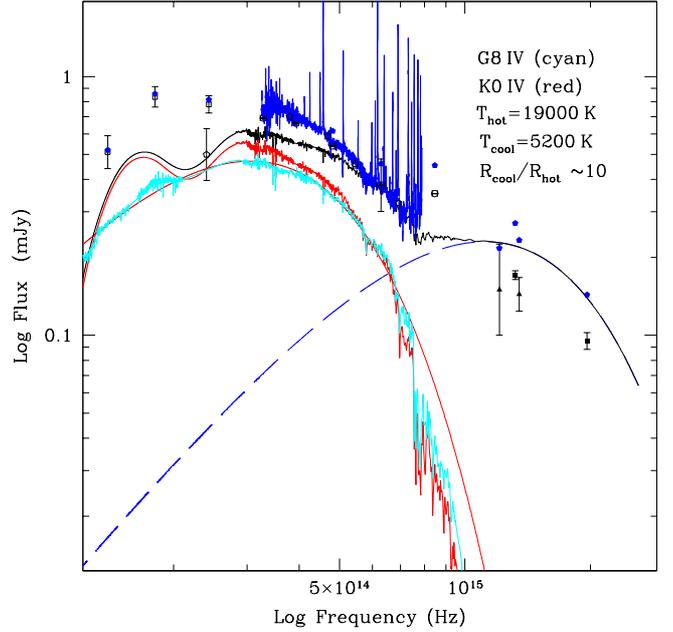}
  \caption{Spectral energy distribution of \sdss\ from UV to IR (right to left). The two black squares are the {\sl GALEX} FUV and NUV measurements,  triangles are {\sl Swift} UVOT  UVM2 \& UVW1 filters   followed by black circles corresponding to the SDSS $ugriz$ bands and our own $BVRI\,J$ measurements marked with open pentagons. Finally, the  three open black squares are 2MASS JHK measurements. The large vertical error bars  are not measurements errors, but show the extent of variability detected in the corresponding filter during long-term monitoring. The filled  blue pentagons are interstellar-extinction-corrected data using E(B-V)=0.056 and the blue dashed line corresponds to SDSS spectrum after de-reddening.  The continuous red and dashed blue lines are simultaneous black body fits to the observations with 5200 and 19000\,K,  respectively. The cool black body corresponds to the presence of  a donor star in the binary system and  is also represented by K0\,IV or G8\,IV spectra of standard stars. The black continuous line is the sum of the K0\,IV spectrum and a 19\,000\,K black body, which appears to fit most of the observed data well. 
  }
  \label{fig:sed}
\end{figure}

The spectral energy distribution (SED) of \sdss\  from the IR to the UV range is shown in Figure\,\ref{fig:sed}. Different black symbols mark photometric measurements from different instruments as indicated in the figure caption.  Since the source is clearly variable,  observations taken at different times do not necessarily match.  The extent of the variability is shown by the bars placed on the data corresponding to the $UVW1 \& UVM2$, $g$ and $J$ bands, where
relatively long continuous observations are available. Therefore these large bars are not measurement errors.

On top of the black symbols, the blue pentagons correspond to the data after a de-reddening
of $E(B-V)=0.056$.  Also shown in blue is the de-reddened SDSS optical spectrum. In section\,\ref{sec:spcl} we have shown that the secondary star is a  late-G /early-K star. The red line in Figure\,\ref{fig:sed} corresponds to a  contribution of K0\,IV star and the cyan line to a G8 subgiant, as was determined in Section\,\ref{sec:spcl}. Although we have argued that the secondary star has a much smaller radius 
than a sub-giant, their 
SED should not differ substantially from our object. For simplicity, the contribution of the secondary can be fitted alternatively by a $\sim 5\,000$\,K black body (smooth continuous red line). Apparently,  the secondary alone with 40-45\% input around 5000\,\AA\ cannot account for the whole IR flux, even if some of the  disparity between the optical and the IR data is caused by the variability of the object.  Moreover, there is a strong unaccounted UV radiation. In theory, the  UV excess can be  fitted with a second black body with a temperature of $\sim 19\,000$\,K  (blue dashed line). The  sum of a K0 star and a 19\,000\,K black body is drawn by a black line.  However,  the origin of the  UV radiation source is not obvious. When  the energy budget is considered, it appears that  the hot component must have a radius only ten times smaller than the cool component  to produce sufficient flux to be represented by the blue dash line in Figure\,\ref{fig:sed}.  This is certainly 
inconsistent with the size of the WD primary.  Thus, the UV excess needs a different explanation than a  black body coming from a stellar component.  A radiation from an accretion disk can also be excluded, because usually there are much higher temperatures in the inner parts of the disk, and the  corresponding SED at these frequencies is a flat power law.  
Fortunately, a SED like this has been observed and successfully interpreted before.  \citet{2006ApJ...646L.147S}, using {\sl GALEX} observations, discovered that the UV flux from the  polar EF\,Eri is much higher than the underlying 9\,500\,K \, 
white dwarf might produce, and that it is highly modulated with the orbital period. Subsequently, \citet{2008ApJ...672..531C} reported IR observations of the same object with SpeX on the {\sl IRTF} and showed that its near-IR SED is dominated by cyclotron emission and  that the cyclotron emission can  satisfactory explain  the UV flux as well. To do this, the white dwarf must have a complex multi-pole structure with a variety of field strengths  instead of a simple centered dipole with only two components.   \citet{2007A&A...463..647B} showed that such complex multi-pole structures exist in more than one object. For EF\,Eri, in addition to low-strength pole components, they inferred a quite large  $B\approx100$\,MG field strength. \citet{2008ApJ...672..531C}  used a $B=115$\,MG  value  to successfully fit the observed UV flux and its variability. We believe that similarly, the IR and UV-excess observed in \sdss\ is caused by  a cyclotron emission from a highly magnetic white dwarf with a complex field structure and a range of strengths  that generates cyclotron lines on both sides of the optical domain.   

\begin{figure}
  \includegraphics[width=8.8cm,  clip]{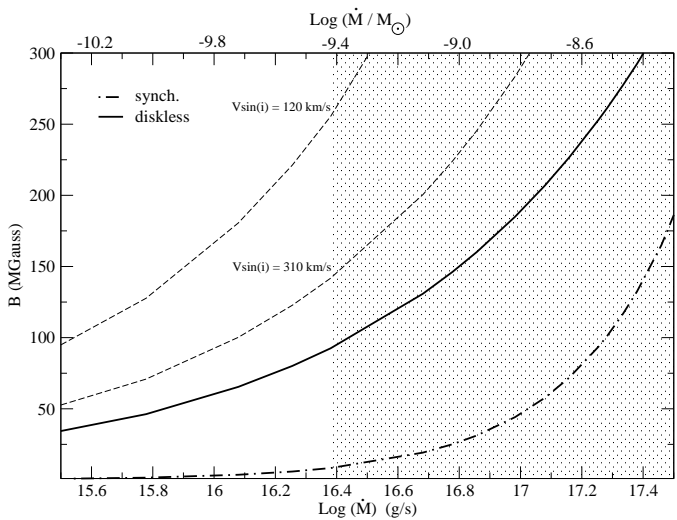}
  \caption{Requirements for synchronizing  the white dwarf spin as a function of the magnetic field of the white dwarf and the mass transfer rate from the donor. The white dwarf spins synchronously if $B$\ exceeds the limit shown by the dash-dotted  line. Above the solid line the accretion disk is absent since  the pressure of the equatorial magnetic field exceeds the free-fall ram pressure of the accretion stream.  The shaded area corresponds to the mass accretion rates determined for \sdss\ from X-ray luminosity.  Isolines of equal velocity calculated for the  treading region, i.e, the highest velocity reached by the stream of  matter from the donor star to the treading radius where the magnetosphere of white dwarf captures matter and diverts it to the white dwarf surface are shown by dashes. This is  the  radial velocity of the emission line components that are expected to be emitted if the stream and, particularly, the coupling region are ionized. }
 \label{fig:BMdot}
\end{figure}

\section{Classification of \sdss}
\label{sec:class}

We  to summarize what we were able to learn about \sdss\ so far.   The object is a binary with a 14.3 hour orbital period. One of the stellar components is a  late-G--early-K  evolved star.  The strong emission lines suggest that it is an interacting binary  with a matter transfer from the visible star to an accreting  compact and undetected component. The highly ionized plasma emits not only in hydrogen and neutral helium lines, but also in He\,{\sc ii}, which is comparable in strength to the intense H$_\beta$.  The radial velocity amplitude and the orbital phasing  of the wings of  the H and He lines advocate that  there is  hot plasma  closely associated with the accreting stellar component. In a classical interacting  close binary system the matter transferred from the donor star to the more massive companion forms an accretion disk, which is the source of emission lines and a hot continuum.  However, an accretion disk hardly ever produces such a strong He\,{\sc ii} line, and even if it does, the emission normally comes from a smaller area with a hot spot. If there is such a spot, the emission lines obtain a narrow component that varies with a different phase than the rest of the disk and is detected as an S-wave within the broader emission lines formed in the bulk of the disk. It is also remarkable is  that the cores of the emission lines are symmetrical, narrow and probably devoid of radial velocity variability. These facts coupled with the flatness of Balmer decrement  speak in favor of more diffuse and static gas than which can be found in an accretion disk.  The form of the SED in the UV also does not correspond to that of an accretion disk. Therefore, we conclude  that the accretion in this object probably occurs not via a disk. 

The alternative is magnetically governed accretion onto the surface of  a highly magnetized white dwarf.     
We propose a polar scenario  to explain the modulated X-ray light curve of the object. It also helps to understand the UV and IR excess in  terms of cyclotron emission from a strongly magnetic white dwarf.  Accordingly, the magnetic nature of this binary system is deduced from  the above mentioned variability and the  strong high-ionization lines of He\,{\sc ii} in the optical spectrum. 
The discovery of  V1309\,Ori (RX\,0515.41+0104.6) at  an eight hour orbital period \citep{1994ApJ...435L.141G}, twice as long as the nearest known long-period polar, caused confusion, because  to maintain the synchronization  of the white dwarf at a separation corresponding to an eight  hour period, the magnetic field must be extremely strong or the accretion rate low, which was not observed  \citep{1995ApJ...443..319S,1996rftu.proc..107B}.  To resolve the paradox,  \citet{1995ApJ...453..446F} suggested that the secondary star in V1309\,Ori, which was identified as an early-M dwarf,  might be slightly evolved, possess a strong magnetic field and stellar wind, which would ensure the necessary accretion rate and synchronization by dipole-dipole locking of two magnetic components. For V1309\,Ori it is difficult to produce an evolved secondary, because the nuclear evolution of an M0--M1 star would take $10^{11}$\,years to expand sufficiently. For  \sdss\  the orbital period is almost twice as  long that of  V1309\,Ori, but the secondary is more massive and the probability that it is evolved is much higher.

We  have no  measurement of the magnetic field, but we can assume that the magnetic field strength is sufficient  to prevent formation of the disk and to synchronize rotation of the white dwarf with the orbital period. We consider the place of \sdss\ in the  \mdot -- $B$ plane (Figure\,\ref{fig:BMdot}). We relied on the prescriptions provided  by \citet{1996Ap&SS.241..263W}, where a modest magnetic moment of the donor star (on the order of 100 G) was also taken into account, to calculate the corresponding relations. The solid line corresponds to the radius of the magnetosphere at which the formation of an accretion disk is prevented  \citep[Eq. 31 in][]{1996Ap&SS.241..263W}. It appears that the condition for synchronization requires significantly weaker fields, marked by the dash-dot line  \citep[ch. 4 in][and references therein]{1996Ap&SS.241..263W}.    With dashed lines we also show the velocity of particles in the ballistic stream when it reaches the coupling region where it is  captured by the magnetosphere and channeled onto the white dwarf.   Assuming that the X-ray luminosity is generated by the accretion on the magnetic pole, we can  estimate  the lower limit of the accretion rate ($\dot{M}\approx10^{16.4}$\,g\,s$^{-1}$).  At  $\log$ \mdot$ > 16.4$  a $B > 80$ MG magnetic field is necessary to prevent  formation of  a disk and at the same time to synchronize the spin of white dwarf. Such a magnetic field will also generate  cyclotron lines in the UV.  At such a radius the ballistic stream has a significant velocity, and if the matter there is ionized, we should see the corresponding velocities in the emission line profiles.

A standard polar scenario  assumes  mass transfer via a stream of matter leaving the L$_1$ point and falling ballistically toward the more massive magnetic white dwarf. The stream of  matter that reaches the magnetosphere of the white dwarf is captured at the so-called treading region and is channeled to the magnetic pole via magnetic lines. In the absence of accretion disk, emission lines of polars originate mostly in the stream (in ballistic and/or magnetic part) and sometimes from  the irradiated face of the donor star. As a result, emission lines in polars usually appear to be multi-component, asymmetric, some with large velocity amplitudes, all modulated with the orbital period with distinct phases.  We see none
 of this in \sdss. If the emission lines originated in the stream, the  wings of the emission lines would be asymmetric and the double-Gaussian procedure would not convolve in most of the orbital phases.  Thus, we assume that mass transfer in this system occurs not through the stream, but rather via stellar wind from a donor star caught in an evolutionary expansion.  We further speculate that 
the bulk of the emission lines is radiated by the ionized wind matter, which is captured by the gravitational and magnetic fields of the  WD  and ultimately channeled onto the magnetic pole of the WD. This channeled part has an intrinsic velocity that broadens the line and is more significant in the wings of the lines. For the same reason this emission line component  becomes narrower when the observer's line  of sight is aligned with  the streamed matter, i.e.,  near the phase when the magnetic pole shines toward the observer and the X-ray flux intensity reaches maximum, as can be seen by comparing Figures\,\ref{fig:rvhem} and  \ref{fig:xuv}.

In a way, this object is  similar to the small group of systems that were called low accretion rate polars (LARP) when they were first discovered \citep{1999A&A...343..157R,2000A&A...358L..45R}, which were recently renamed pre-polars \citep{2005ApJ...630.1037S}. \citet{2011A&A...530A.117V} argued that the magnetic WD in these  CVs captures matter from the weak wind ($10^{-13}$\msun yr$^{-1}$) of pre-main sequence red dwarfs. Others believe there is a stream departing from  L$_1$  \citep{2007A&A...468..643T,2010MNRAS.403..755K}.  
In \sdss\,  however, the mass accretion rate  is at least $10^3$ times higher than in pre-polars that contain M-type donors and 100 times higher than the wind of a main-sequence mid-G star can provide. Nuclear evolution of the donor to the terminal age main-sequence will not increase mass loss rate by the wind  to such a degree. In a close binary with coupled magnetic fields the wind-loss process  can be  different from non-magnetic systems, but the angular momentum loss rate  differs from the  ``standard'' one that is  usually applied in CV studies by a factor of several only \citep[e. g.,][]{2012ApJ...746L...3C}.  There are  no specific models  to describe the situation with  \sdss\, and our observations do not provide sufficient information to unambiguously identify the mode of mass transfer. It might be possible that the donor star fills its Roche lobe, but we are not able to see the stream  because of the orientation of the flow, lack of ionization of the high-velocity component, or maybe because the interacting magnetic fields prevent ballistic trajectory. 

Despite the possible difference in evolution, in the absence of any more realistic models, 
we  illustrate the possible origin of \sdss\ with the model for a  non-magnetic system  in which the donor is 
filling its Roche lobe and losing mass via $L_1$. 
We  speculate that  whatever angular momentum loss mechanism brought the donor star into  contact, 
the dominant factor that determines evolution of the system after
Roche lobe overflow (RLOF) by the donor is the evolutionary state of the latter
at the instant of overflow.  Then, if the time scale of angular
momentum loss by the system is not drastically different from the time scale of that process 
in the ``standard'' scenarios applied in the studies of non-magnetic 
CVs, as concluded by \citet{2012ApJ...746L...3C}, we may expect that  the subsequent
evolution will be close to the path presented below.
We also expect that in the RLOF mode of mass transfer the magnetic nature of the white dwarf would not 
alter the evolutionary scenario developed for non-magnetic systems. 

\begin{figure}
  \includegraphics[width=9cm,  clip]{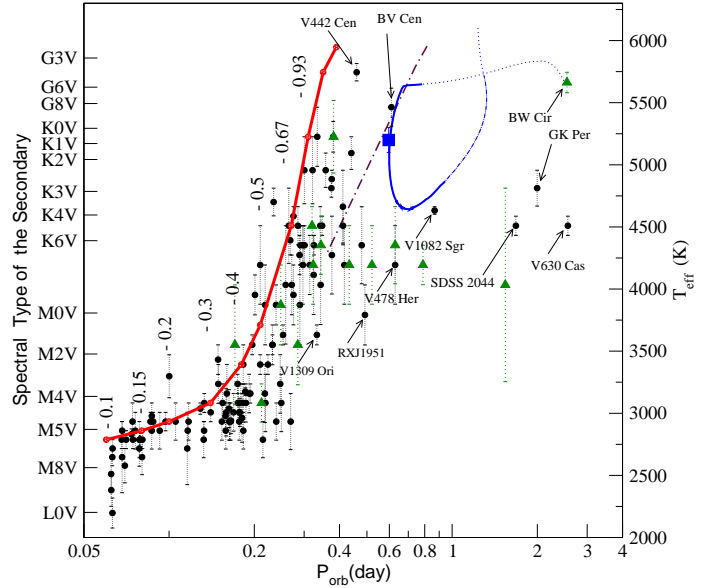}
  \caption{Spectral types and effective temperatures of the donors of CVs (black dots) and LMXB (green triangles) plotted vs. their orbital  periods (an update of the 
  \citet{1998A&A...339..518B} list).  CVs with remarkably long periods are marked individually. 
 \sdss\ is shown by a solid blue square. The black dot-dashed line marks the  bifurcation period.
The red line corresponds to the periods of semi-detached systems with main-sequence donors and 0.7\,\msun\ accretors. 
The evolutionary track of the system with initial masses of donor and accretor equal to 1\,\msun\ is plotted with the 
blue line. Dotted parts of the line correspond to a detached state, solid parts of the line  to the semi-detached state, while the dash-dotted part corresponds to the semi-detached state in which the mass transfer rate is below 
$10^{-10}$\,\msun\,yr$^{-1}$ and the system may  hardly be observed. 
  (i) effective temperatures of CVs below about 4500\,K (later than K5) are considered as unreliable and (ii) evolutionary models with gray atmosphere
boundary conditions overestimate  the \te\ of low-mass stars \citep{1997A&A...327.1039C}. 
}
  \label{fig:limits}
\end{figure}

In Fig.~\ref{fig:limits} we present  an orbital period -- spectral type of the donor -- effective temperature  diagram for CVs.  
The thick red line connecting  dots for different mass stars shows this relation based on Eq.~
(2.87) from \citet{1995CAS....28.....W},  which assumes  that the secondaries follow the empirical mass-radius 
relation for main-sequence stars;  the calibration of spectral  types and effective temperatures is  based on 
empirical data from \citet{2006MNRAS.373..484K} and \citet{2007MNRAS.382.1073M}.
Systems with low-mass donors mostly remain close to the red line since the donors are not nuclearly evolved. 
Their deviation from this line is discussed in detail by
\citet{2011ApJS..194...28K}. The donors  with initial mass $0.95 \aplt M_{2}/M_\odot \aplt 1.3$  may exhaust a significant fraction of the hydrogen in their cores  prior to RLOF. Then, depending on the central hydrogen abundance  
 of the donor ($X_c$), upon RLOF,  newborn CV might evolve to shorter or longer 
orbital periods \citep{1985SvAL...11...52T,1987SvAL...13..328T}, i. e.,
 ``converge'' or to ``diverge''  (in the terms coined by  \citet{1988A&A...191...57P}).
For every mass combination of components in a pre-CV there  is a certain critical value 
of the orbital period at RLOF that divides binaries into converging and diverging ones 
(''bifurcation period'', $P_{\rm b}$). 
 Bifurcation periods calculated for systems with 0.6\,\msun\ accretors and (0.95 -- 1.3)\,\msun\ 
donors  are shown in Fig.~\ref{fig:limits} with a dash-dotted line.
Systems in which RLOF occurs at periods only slightly shorter than $P_b$ (corresponding to  $X_c \aplt 0.1$) evolve into ultracompact binaries with \porb $\aplt$ 80 min., while the systems with only
slightly longer periods at RLOF (initial masses of He-cores $\lesssim0.01$\,\msun) meander
close to  $P_{\rm b}$\ before evolving to long \porb. These systems may spend $\apgt $100\,Myr at periods $\sim$1 day, i. e., still  in the CV-range.

We show in Fig.~\ref{fig:limits} the  track of a system with initial masses of donor and accretor
 1\,\msun,  computed  assuming that angular momentum loss follows the \citet{2011ApJS..194...28K} semi-empirical  law
(Yungelson et al. {\it in prep.})\footnote{For computations, an appropriately modified version of P.P. Eggleton 
code  \citep{1971MNRAS.151..351E,1995MNRAS.274..964P} and {\it priv. comm. 2006}, was used.}. 
In the semi-detached stage, the mass of the accretor was kept constant, since expected mass-exchange rate is low and it is assumed that all matter transferred to the donor is, furthermore,
lost from the system due to hydrogen-burning shell  explosions that reduce specific angular momentum of the accretor. 
The initial (post-common-envelope) period of the system is 2.5 days. The track crosses the position of 
\sdss. 
The donor fills its Roche lobe at \porb $\approx0.68$\,day, when hydrogen  just became 
exhausted in its center ($X_c\approx 10^{-12}$).  
At the position of \sdss, which is reached in $\approx 80$\,Myr after RLOF,
the mass of the donor is close to 0.8\,\msun, complying with the estimates of the donor mass of \sdss, and  
the mass-loss rate is $1.7 \cdot 10^{-9}$\,\msun\,yr$^{-1}$. 
In the course of the preceding evolution  $\dot{M}$ was close to this value; 
in our evolutionary picture this implies that \sdss\ may be a former nova.  
The accretion efficiency  of the stream matter in non-magnetic CV is estimated to be 20 -- 30 per cent
\citep{2012PhyU...55..115Z}. Then,  using 
expression $L_{\rm X} \approx (1/3)GM_a\dot{M}_a/R_a$ for the X-ray luminosity of CVs 
in the (0.2-10)KeV range 
and $R_a=10^9$cm, we obtain $L_{\rm X}\simeq 10^{33}$ erg s$^{-1}$, in agreement with observations. The mass transfer rate  drops below $10^{-10}$\,\msun\,yr$^{-1}$ shortly after orbital period turnaround  toward longer periods when the  system, probably, will become hardly detectable (at least, if it is non-magnetic)\footnote{In this respect,
it will be similar to the post-period-minimum ordinary CVs.}. 
It exists as a CV for $\approx$2.7\,Gyr. The final state of the system is a detached pair of white dwarfs.

\section{Conclusions}

We presented multi-wavelength observations 
 of the close binary system \sdss. The orbital period of the system is 
0.5941 days based on the radial velocity variability of the absorption lines emanating from the  late-G/early-K donor star. The radial velocity of the emission lines, the X-ray and UV radiation  were all modulated with the  period corresponding to the orbital one. We argued that the compact binary contains a highly magnetic accreting white dwarf with a spin period synchronized to the orbital period, which produces X-rays and cyclotron emission. The latter is not detected directly but was invoked to explain the variable IR and UV excess radiation. This makes \sdss\  the polar with the longest known orbital period. However, several key questions remain unresolved. The magnetic field does not manifest itself directly and we only suspect that it has a complex multipole structure with low and high field strengths.  We are also not sure how the mass transfer proceeds. Roche-lobe overflow is required for the mass-transfer rate   $\sim10^{-9}$\,\msun\,yr$^{-1}$, which may sustain  an X-ray luminosity of  $\sim 10^{33}$ erg s$^{-1}$, as estimated based on the shortest  distance that  assumes a main-sequence donor-star. 
Meanwhile, we do not have any observational evidence of the  stream  common for CVs.
But we  presented  firm evidence that the donor star has to be significantly evolved and larger than a normal main-sequence star of corresponding spectral type. 
This provokes speculation that maybe the magnetic fields of the compact star and evolved donor are coupled (which is also required  to keep the rotation of the white dwarf  synchrone with the orbital period at such a long period/separation)  and the mass transfer proceeds quite differently. 
The models presented by \citet{2012ApJ...746L...3C}, while being  far from  realistic, indicate that the white dwarf magnetic field results in a significant distortion of the companion field  in  ways that differ according to dipole alignments, white dwarf field strength, and orbital separation. It was shown that  strong magnetic fields of the white dwarf could inhibit the donor wind outflow, reducing open field regions and instead capturing the stellar  wind from the donor and reducing the magnetic braking relative to single stars. Whether this might provide a sufficient mass accretion rate to fuel the observed X-ray luminosity is still doubtful. However, if that happens 
and the systemic angular momentum loss time scale is strongly different from the one according to the    \citet{2011ApJS..194...28K}  law,  the evolutionary scenario for \sdss\ may be  different from the one presented in Section\,\ref{sec:class}. 
If the donor star fills its Roche lobe and transfers matter in the  manner as other CVs,  we can expect the system to evolve into a wide pair of white dwarfs.  We emphasize that  other paths of evolution, if existent,  are very intriguing, considering the fact that \sdss\ contains quite heavy white dwarf  already close to the Chandrasekhar limit. 

Further studies are required to clarify the nature of this object and validate speculations implied in this paper.  A new  time resolved UV and IR observations are vitally important to confirm the presence of cyclotron lines and  affirm claims of magnetic nature of the accretion in this interesting binary system. 

\section*{Acknowledgments}

DGB is grateful to CONACyT for providing  the financing  his doctoral studies and to the organizers of the IAU Symposium 281 for a travel grant to attend the conference.  
GT and  SZ acknowledge PAPIIT grants IN-109209/IN-103912  and CONACyT grants 34521-E; 151858 for resources provided for this research. JE has been supported by the  PAPIIT grant IN122409.
 LRY is indebted to P.P. Eggleton for providing a copy of his evolutionary code. 
LRY is supported by RFBR grant
10-02-00231 and the Program of the Presidium of Russian academy of
sciences P-21. 

\bibliographystyle{aa} 

\end{document}